\documentclass[twocolumn,showpacs,preprintnumbers,amsmath,amssymb]{revtex4}
\usepackage{graphicx}
\usepackage{dcolumn}
\usepackage{bm}
\begin{document}
\preprint{APS/123-QED}
\title{HOW TO CONSTRUCT OF MECHANICS OF THE STRUCTURED PARTICLES }
\author{V.M. Somsikov}
 \altaffiliation[] {}
 \email{vmsoms@rambler.ru .
 Site: http://sites.google.com/site/peosrussian/}
\affiliation{%
Laboratory of Physics of the geoheliocosmic relation, Institute of
Ionosphere, Almaty, Kazahstan.
}%

\date{\today}
\begin{abstract}
The mechanics of structured particles ($SP$) consisting from
potentially interacting material points are discussed. For this
purpose the derivation of the $SP$ equation motion in the field of
external forces is submitted. The differences between the
properties of the dynamics of material point and properties of the
dynamics of $SP$ are analyzed. The explanation of how mechanics of
the $SP$ leads to the account of dissipative forces is submitted.
The derivation of Lagrange, Hamilton and Liouville equations for
$SP$ Is shown. The question of why the motions of the $SP$ are
determined by the two types of symmetry: internal symmetry of the
system and symmetry of space and how it leads to two types of
energy and forces accordingly are discussed. It is shown how the
concept of entropy arises in classical mechanics. The question how
the mechanics of the $SP$ leads to thermodynamics, statistical
physics and kinetics is explained.
\end{abstract}

\maketitle

\section{\label{sec:level1}Introduction\protect}

Structure of the world is hierarchical in nature. Upper stage is
the Universe. It is composed from galaxies. Galaxy, in its turn,
is composed of structural elements. At the lower hierarchical
level are molecules and atoms. But they are also systems
consisting of "elementary" particles. It is hard to tell how deep
it goes down the hierarchical ladder. The limit of divisibility of
matter has not yet been found. Thus the world is a hierarchy of
systems. Therefore, we can say with confidence that the systems
lie at the basis of the world, rather than elements. From this it
follows that for correctly building a picture of the world we need
to know the laws of creation, interaction and evolution for the
systems' but not for indivisible elements as it have a place in
modern physics.

Since all the bodies have a structure, they have different
properties than the elements. First of all, they have internal
energy which appeared due to the relative motions of their
elements. Therefore as a result of motion of bodies in the
external field of forces some part of the energy will be go into
the internal energy. Therefore the trajectories of motion of real
bodies are defined not only by the change of the motion energy but
by the change of the internal energy also.

The Newton's motion equation is constructed based on the model of
structureless bodies. Therefore it does not include the terms that
are responsible for energy which increases the internal energy and
dissipated in the environment [1-4]. In practice, this part of the
energy is taken into account in an empirical manner by the
addition of friction forces to Newton's motion equation. The
friction work determines the dissipative part of the motion
energy. The coefficient of friction, which determines a
transformation of body's motion energy into internal energy, is
taken from experiment. So the second Newton law does not account
for structurality of real bodies. To eliminate this shortcoming it
is necessary to change the unstructured body model on the model of
bodies consisting from elements. The motion equation and laws of
interaction of these bodies must be determined from the conditions
of implementation of Newton's laws for the elements.

In the local equilibrium approximation any non-equilibrium system
can be represented by a set of equilibrium system
consisting of a sufficiently large number of potentially
interacting material points ($MP$) which has a relative motion to
each other [10, 12]. In the thermodynamic limit at enough weak
interactions, each of the ($SP$) can be regarded as equilibrium
during the entire process. Consequently, the properties of their
dynamics can be studied using the $SP$ motion equation since it
describes the process of equilibration.

Thus the mechanics of $SP$ based on
the known dynamic properties of the $MP$ must be constructed to investigation of the non-equilibrium systems. The $SP$ mechanics can be constructed on the basis of Newton's
laws for $MP$  in the frames of the following restrictions [5-8]:

1). Every $MP$ belongs to its $SP$ during all process.

2). $SP$ is in equilibrium during all time.

The first restriction eliminates inessential complications related
to the necessity to including of $MP$ transition between $SP$.

The second restriction is equivalent to the requirement of weak
interaction.

The aim of this paper is to show how to construct mechanics of
$SP$, based on Newton's laws for $MP$ and what are the qualitative
differences between the dynamics of $SP$ from the dynamics of
$MP$. For this purpose mainly will explained the following
questions: how the motion equation for $SP$ can be obtained from
the symmetry properties of the space and time; how based on the
$SP$ motion equation possible to derive the laws of
thermodynamics, statistical physics and kinetics; how the concept
of entropy can be introduced in the classical mechanics; what is a
nature of the deterministic irreversibility.

\section{THE SYMMETRY PROPERTIES OF $SP$ MECHANICS}
Let us explain how the $SP$ motion equation can be obtained based
on the symmetry properties space and systems.

If we have a $SP$ where interactions between $MP$ are absent, the
motion equation of $SP$ is the sum of the independent $MP$ motion
equations. In this case the motion of $SP$ is determined by the
sum of solutions of the $MP$ motion equations. But the presence of
$MP$ interactions precludes the summation of equations of motion.
Thus the presence of interactions between the $MP$ systems is
equivalent to the imposition of additional restrictions or
external links [3]. On the example of the two-body problem it can
be shown that the interaction between $MP$ leads to the
interdependence of the coordinates and velocities of $MP$ in the
laboratory coordinate system ($LCS$) [7]. Hence it is clear that
the symmetry of the motion equation for the system will be
different than the symmetry of Newton's equation for each $MP$.

It is well known that for solving the tasks $N$- body systems
necessary to find such transformation of the coordinate of system
which leads to separation of the variables. It is equivalent to
find transformation of the vector space $L_q$ of the dimension $q$
which determined the motion of the system,  into another vector
space of the same dimension but in which the vector space splits
into independent orthogonal subspaces. In the language of group
analysis [9] it is equivalent to the separation of the group
representations $T(G_a)$ of the symmetry of the equations of
motion of the $G_a$ group (here $a=1,2,3...n$ -number of elements
in the $G_a$ group of symmetry) in the vector space $L_q$ into
irreducible representations $T_i\subset{T_q}$, where $i=1,2,3...k$
and $k$-is a number of irreducible representations of the group of
symmetry, where $T_q=T_1\oplus{T_2}...\oplus{T_k}$. Each
irreducible representation $T_i$ acted in subspace $L_i$, for
which we can write $L_q=L_1+L_2...+L_k$. Thus the vector
$r_q=\sum\limits_{i=1}^{k}{r_i}$ in $L_q$ space decomposed into
irreducible components $r_i$. If $q=k$, i.e. $L_q$ decomposes into
the basis vectors, the system's motion equation is integrable. It
turns out that for the $MP$ system there exists a transformation
of the vector space where it splits into two orthogonal subspaces.
Thus we have: $L_q=L_{ins}+L_{out}$, where $L_{ins}$ is a subspace
which determined by the internal degree of freedom, $L_{out}$ is a
subspace which determined by the symmetry of space. Such
replacement of coordinate system corresponds transition to the
coordinate system in which the motion of the system decomposes
into the motion of center of mass ($CM$) and the motion of $MP$
with respect to the $CM$ system.

The symmetry of the system is determined by its structure and the
nature of the interactions between $MP$. The $MP$
interactions is determined by the distance between them.
Therefore the distance between $MP$ should be choose as the
independent variables. In our case these variables will determine the internal
dynamics of the system. As the variables that determine the motion
of the system in space, we should take the coordinates of the $CM$
of the system. As it easy to see on the example of two-body
problem, this a set of variables forms a complete basis of the
independent variables that determined the system's dynamics in
space [7]. In the next section will shows why and how in these variables
the system's energy divided into energy of motion of the system as
a whole and the internal energy of the system.

In the new coordinate system we have the motion of the $CM$ in
space and motion of the $MP$ relative to the system's $CM$. The
motion of the $CM$ system is determined by macro variables
(coordinates and velocity of the $CM$ system). The motion of $MP$
with respect to the $CM$ expressed in terms of micro variables.
Thus the vector space in which the $SP$ motion is defined splits
into two orthogonal subspaces from macro and micro variables [7].

The motion of an element that has no internal structure in the
external field of forces is determined only by the symmetry of the
space. But the motion of the system is determined by two types of
symmetry: the symmetry of space and symmetry of the system. In
this case according to the Noether's theorem, the dynamics of the
system is determined by two types of energy. This is the internal
energy and the energy of the system. These energies are
independent. They correspond for two types of forces that
determine the dynamics of the system. One type of forces
determines the system motion; the other is responsible for the
$SP$ internal energy.

The forces that change the internal energy
do not change the $SP$ velocity because their sum is zero. These
forces can only be found through the value of their work.
Therefore the $SP$ motion equation can be determined from the
energy of the system. This approach allows us to get all the
collective forces that determine the dynamics of $SP$, without
imposing on them the requirements of potentiality, as is done
in obtaining of the principle of least action [2, 3]. Thus we
will construct the mechanics of $SP$ based on the energy.

\section{THE ENERGY OF $SP$}
Let us define an expression for the energy of a system of
potentially interacting $MP$ with unit mass, i.e. $m=1$. In a
homogeneous space the energy of such a system is not changed
although the energy of each $MP$ may vary due to their
interactions. Therefore in a homogeneous space the momentum of
$SP$ is conserved. Its preservation means the constancy of the
$CM$ velocity. The $CM$ velocity is equal to:
$V_N=\dot{R}_N=(1/N)\sum\limits_{i=1}^{N}{\dot{r}_i}$, where
${\dot{r}_i}$- velocity for $i$-$MP$. From here the kinetic energy
of the system motion is equal to $T_N^{tr}=M_NV_N^2/2$, where
$M_N=Nm$. This energy coincides with the energy of the body with
mass $M_N$ which moving with the velocity of $CM$.

Let us show that the internal kinetic energy $T_N^{ins}$ is equal
to the sum of the kinetic energies of $MP$ with respect to the
$CM$.

From equality  $N\sum\limits_{i=1}^{N}
{v_i}^2={V_N^2+\sum\limits_{i=1}^{N-1}\sum\limits_{j=i+1}^{N}v_{ij}^2}$
is following that $T_N=
{M_N}V_N^2/2+{m/N}{\sum\limits_{i=1}^{N-1}\sum\limits_{j=i+1}^{N}v_{ij}^2}/2$
(a), where $v_{ij}=v_i-v_j$. The first term in (a) is $T_N^{tr}$.
Since the total kinetic energy of the system is equal to the sum
of the kinetic energy of the $CM$ and the kinetic component of the
internal energy. Then we have:
$T_N^{ins}={m/N}{\sum\limits_{i=1}^{N-1}\sum\limits_{j=i+1}^{N}v_{ij}^2}/2$.

Let us transform the energy by a change of variables:
$v_i=V_N+\tilde{v}_i$, where $\tilde{v}_i$ is a $MP$  velocity
relative to the $CM$. We obtain:
$T_N=M_NV_N^2/2+\sum\limits_{i=1}^{N} m{\tilde{v}_i}^2/2$ because
$\sum\limits_{i=1}^{N}{\tilde{v}_i}=0$.

In according with (a) we have: $\sum\limits_{i=1}^{N}
m{\tilde{v}_i}^2/2={1/N}\sum\limits_{i=1}^{N-1}\sum\limits_{j=i+1}^{N}mv_{ij}^2/2$.
Hence the kinetic energy of the relative motion of the $MP$ is
equal to the sum of the kinetic energy of motion relative to the
$CM$ and the total kinetic energy of the system is
$T_N=T_N^{tr}+T_N^{ins}$.

Because $r_{ij}=\tilde{r}_{ij}=\tilde{r}_{i}-\tilde{r}_{j}$, where
- $\tilde{r}_{i},\tilde{r}_{j}$ is a coordinates of $MP$ relative
to the $CM$ then the potential energy of $i$ and $j$ $MP$
interaction is $U_(r_{ij})=U_(\tilde{r}_{ij})$. From here the
potential internal energy is equal to
${U_N}={\sum\limits_{i=1}^{N-1}}{\sum\limits_{j=i+1}^{N}}U_{ij}(r_{ij})
$, and total internal energy of $SP$ is $E_N^{ins}=T_N^{ins}+U_N$.
In homogeneous space the energies $T_N^{ins}$ and $U_N$ are
invariants of the motion.

Because $\sum\limits_{i=1}^{N}{\tilde{v}_i}=0$ then we
have:$\sum\limits_{i=1}^{N}{\dot{\tilde{v}}}_i=0$ . It means that
the sum of the internal forces is equal to zero. But this sum is
the sum of the forces of $MP$ interaction. So the internal forces
can't change the momentum of the $CM$. Therefore micro and macro
variables are independent. If the system does not affect the
properties of space then its total energy is the sum of the
kinetic energies of the $MP$, the potential energy of mutual
interaction and the potential energy determined by the
inhomogeneity of the space. I.e. $E_N=T_N+U_N+U^{env}=const$.
Separating the internal energy, we can write:
\begin{eqnarray}
E_N=T_N^{tr}+E_N^{ins}+U^{env}, \label{eqn1}
\end{eqnarray}
where $E_N^{ins}=T_N^{ins}+U_N$, is internal energy,
$T_N^{ins}=\sum\limits_{i=1}^{N}m\tilde{v}_i^2/2$ is a kinetic
part of internal energy, $U_N$ is a potential part of internal
energy, determined by the interactions of $MP$.

Thus the internal energy and the energy of motion along the
trajectory of the $CM$ system changes so that their sum is always
constant. It is the law of conservation of energy of the systems.

\section{THE MOTION EQUATION OF $SP$}
Differentiating the energy of $SP$ with respect to time, we obtain
[7, 8]:
\begin{eqnarray}
V_NM_N\dot{V}_N+{\dot E}_N^{ins}=-V_NF^{env}-\Phi^{env}\label{eqn2}
\end{eqnarray}
Here $F^{env}=\sum\limits_{i=1}^{N}F_i^{env}(R_N,\tilde{r}_i)$;
${\dot E}_N^{ins}={\dot T}_N^{ins}(\tilde{v}_i)+{\dot
U}_N^{ins}(\tilde{r}_i)$=
$\sum\limits_{i=1}^{N}\tilde{v}_i(m\dot{\tilde{v}}_i+F(\tilde{r})_i)$;
 $\Phi^{env}=\sum\limits_{i=1}^{N}\tilde{v}_iF_i^{env}(R_N,\tilde{r}_i)$;
$r_i=R_N+\tilde{r}_i$; $M_N=mN$; $v_i=V_N+\tilde{v}_i$;
$F_i^{env}(R_N,\tilde{r}_i)$-is external force which acts on the
$i$-th $MP$; $\tilde{r}_i$, $\tilde{v}_i$ are the coordinates and
velocity of $i$-th $MP$ in the $CM$ system; $R_N,V_N$ are the
coordinates and velocity of the $CM$ system.

Eq. (2) represents the balance of system energy in the field of
external forces. The first term in the left hand side gives the
change in kinetic energy of the system. The second term determines
the change of its internal energy. Thus in the micro and macro
variables the work of external forces splits into two terms.

Now let us take the external forces which scale of heterogeneity
is commensurable with the systems scales. In this case we can
write:$F^{env}=F^{env}(R+\tilde{r}_i)$ where $R$ is the distance
from the source of force to the $CM$ of the system. Let us assume
that $R>>\tilde{r}_i$. In this case the force $F^{env}$ can be
expanded with respect to a small parameter. Leaving in the
expansion terms of zero and first order we can write:
$F_i^{env}=F_i^{env}|_{R}+(\nabla{F_i^{env}})|_{R}\tilde{r}_i$.
Taking into account that $\sum\limits_{i=1}^{N}\tilde{v}_i
=\sum\limits_{i=1}^{N}\tilde{r}_i=0$ and
$\sum\limits_{i=1}^{N}F_{i}^{env}|_{R}=NF_{i}^{env}|_{R}=F_0^{env}$,
we get from (2):
\begin{eqnarray}
V_N(M_N\dot{V}_N)+
\sum\limits_{i=1}^{N}\tilde{v}_i(m\dot{\tilde{v}}_i+F(\tilde{r})_i)\approx\nonumber\\\approx
-V_NF_0^{env}-\sum\limits_{i=1}^{N}({\nabla}F^{env}_{i}|_{R})\tilde{v}_i\tilde{r}_i\label{eqn3}
\end{eqnarray}

In the right-hand side of equation (2) the force $F_0^{env}$ in
the first term depends on $R$. It is a potential force. The second
term depending on coordinates of $MP$ and their velocities
relative to the $CM$ of the system determines changes in the
internal energy of the system. It is proportional to the
divergence of the external force. Therefore, in spite of the
condition $R>>\tilde{r}_i$ the values of $\tilde{v}_i$ may be not
small, and the second term cannot be omitted. Forces corresponding
to this term are not potential forces since we can not express
them in terms of a gradient of a scalar function. So the change of
the internal energy will not equal to zero when the characteristic
scale of inhomogeneities of the external field is commensurable
with the scale of the system.

Multiplying eq.(2) by $V_N$ and dividing by $V_N^2$ we find the
equation of a system motion [5-8]:
\begin{eqnarray}
M_N\dot{V}_N= -F^{env}-{\alpha_N}V_N\label{eqn4}
\end{eqnarray}

where $\alpha_N=[{\dot E}_N^{ins}+\Phi^{env}]/V_N^2$  is a
coefficient determined by the change of internal energy.

Unlike the Newton's motion equations, in the right hand side of
eq. (4) the additional term which determines the change in
internal energy is appeared. This term depends on the time.
Therefore the symmetry of the time of the motion equation for the
systems is another  then the symmetry of the time for the Newton
motion equation for $MP$.

The equations of Aristotle and Newton are special cases of
equation (4). Indeed, according to Aristotle, the motion equation
has the form: $\alpha_NV=F$. As can be seen from (4) this
condition occurs when the friction force is equal to the external
force and acceleration of the system is zero. At the beginning of
the movement by the friction force for system can be neglected.
Then we obtain Newton equation: $m{\dot{v}}=-F$.

The state of this system composed from a set of $SP$ can be
defined in the phase space which consists of $6R-1$ coordinates
and momentums of $SP$, where $R$ is a number of $SP$. Location of
each $SP$ is given by three coordinates and their moments. Let us
call this space as $S$-space for $SP$ in order to distinguish it
from the usual phase space for $MP$. The $S$-space unlike usual
phase space is compressible though total energy of all $MP$ is a
constant. It is caused by transformation of the motion energy of
$SP$ into their internal energy. The velocity of the $CM$ and the
internal energy change when the system moves from one point of
$S$-space to another. Therefore, the system does not return to
initial state if you rotate the velocity of all $SP$. This
ambiguity does not exist in the usual phase space which defines
the position of all $MP$. $S$-space coincides with the usual phase
space when the internal energy of the $SP$ does not change.

Hamiltonian formalism is constructed based on the Newton's motion
equation [8]. But the Newton's motion equation is followed from
eq. (4) when the change of internal energy is absent. Therefore,
the Hamiltonian systems can be regarded as a special case of
dissipative systems.

Because the dynamics of systems from $SP$
characterized by the trajectory of the $CM$ in the $S$-space, the
class of dissipative systems is convenient to call as $S$-systems.
Most likely, $S$-systems as well as the Hamiltonian's systems are
the subsystems of a larger class which is still to determine. This
follows from the fact that the $SP$ model are simplification of
real bodies, although more appropriate for reality than the $MP$
model.

The fact that we're able to determine the $SP$ motion equation
from expression for its energy or Hamiltonian; follows from the
validity of homogeneity of time condition, both for $MP$ and for
the $SP$ as a whole. The internal energy is expressed through the
micro variables which form vector space independent with respect
to macro variables space in which the $SP$ motion determined.

The internal energy can't be transformed into the system's motion
energy. It is follows from the momentum conservation law of the
system. The change of the internal energy is going due to the
non-potential collective forces. These forces are changing the
$MP$ velocities with respect to the $CM$ but they can't increase
the $SP$ velocity. At the same time a potential component of the
total external force that determines the rate of change of the
$CM$ does not change the internal energy. An important fact is
that the forces that change the internal energy are determined by
the nonlinear terms. These terms depend on the gradient of the
external forces [6].

\section{The $SP$ mechanics and thermodynamics}

The task of the thermodynamics is the description of the dynamical
processes in the systems consisting from the large number of
elements [10]. The thermodynamic method of describing the system
is inherently phenomenological. This method does not explain the
physical laws of processes in the systems but reveals the
characteristic relation between the parameters that determine the
systems' state: the temperature, pressure, density, entropy, etc.
The fact that thermodynamics does not allow to understand the
physics of the processes is its biggest weakness. To eliminate
this drawback should find a connection the laws of classical
mechanics with thermodynamics laws. That is the laws of
thermodynamics should be follow from the laws of classical
mechanics. Let us explain how this problem can be solved with the
help of the mechanics of $SP$ [8, 10].

The work of external forces in thermodynamics breaks up into two
parts. One part is related to the reversible work. Another part of
energy goes into heating system. According to it, the basic
equation of thermodynamics looks like [10]:
\begin{eqnarray}
dE=dQ-PdY\label{eqn5}
\end{eqnarray}

Here $E$ is the energy of a system; $Q$ is the thermal energy; $P$
is the pressure; $Y$ is the volume. As we deal with equilibrium
systems, then $dQ=TdS$, where $T$ - temperature, $S$ - entropy.

According to the eq. (5), coming into the system energy can be
divided on two parts. There are energy of relative motion of the
$SP$ and its internal energy. It was showed [5] that in
thermodynamics to the change the $SP$ energy of relative motion
corresponds to $PdV$, and the change in $SP$s internal energy
corresponds to $TdS$. Thus, we will come to the basic
thermodynamic equation if one carries out standard transition to
thermodynamic parameters in the equation (4) [6,7].

Let us take the system consisting from "$R$" numbers of $SP$. Each
$SP$ consists from $N_L$ number of $MP$ and $N_L>>1$, where
$L=1,2,3,...R$ , $N=\sum\limits_{L=1}^{R}N_L$. Then the share of
energy, which goes on internal energy increasing, is determined by
the expression [5-7]:
\begin{equation}
{{\Delta{S}}={\sum\limits_{L=1}^R{\{{N_L}
\sum\limits_{k=1}^{N_L}\int[{\sum\limits_s{{F^{L}_{ks}}v_k}/{E^{L}}]{dt}}\}}}}\label{eqn6}
\end{equation}
Here ${E^{L}}$ is the kinetic energy of $L$-$SP$; $N_L$ is the
number of elements in $L$-$SP$; $L=1,2,3...R$; ${R}$ is the number
of $SP$; ${s}$ is the number of external elements which interact
with ${k}$ element belonging to the $L$-$SP$; ${F_{ks}^{L}}$ is
the force acting on the $k$-element; $v_k$ is the velocity of the
$k$- element.

The eq. (6) can be viewed as an entropy definition in the
classical mechanics. This definition of the entropy corresponds to
Clausius one [10, 12]. The only difference is that classical
entropy follows from analytical expression for the change of an
internal energy obtained by us on the basis of Newton's laws. Thus
the internal energy is the energy of the chaos.

From the Eq. (6), it is possible to obtain the value of the
entropy production and obtain the conditions which necessary to
sustain the non-equilibrium system in the stationary state [10].

Mechanics of $SP$ leads to statistical physics and kinetics.
Indeed, the velocities of $SP$ are determined by average values of
velocities of $MP$. The sum of the $MP$ velocities relative to the
$CM$ of $SP$ is equal to zero. It means that the parameters which
defining the dynamics of $SP$ can be express through the first and
second moments of $MP$ function of distribution [12].

\section{Deterministic irreversibility}

The mixing is inherent property of the Hamiltonians systems. This
property and the "coarse-grain" of the phase space hypothesis are
used in the basis of the currently known explanations of
irreversibility which we will call probabilistic irreversibility
[11].

The "coarse-grain" hypothesis is equivalent to postulating the
existence of fluctuations of the external restrictions on the
system. But this hypothesis contradicts to the determinism of
classical mechanics because it is inconsistent with the laws of
classical mechanics. This disadvantage is eliminated in submitted
here the deterministic mechanism of the irreversibility [8]. Below
we briefly explain the nature of the deterministic irreversibility
and compare it with the explanation of probabilistic
irreversibility.

Key questions which related to the problem of deterministic
irreversibility are the next questions: why irreversibility for
the $SP$ followed from reversibility of the Newton's motion
equation for one $MP$; how this irreversibility from the Newton's
laws for $MP$ is followed; what are the limitations of classical
mechanics which do not allow to come to the deterministic
irreversibility for to one $MP$.

The motion equation for $SP$ has been obtained subject to the
fulfillment the Newton's laws for $MP$. Indeed, the energy of $SP$
is the sum of the energies of $MP$. The energy of each $MP$ is
equal to the energy of motion and potential energy in the external
field and the field of forces acted from another of $MP$. The
motion equations an each $MP$ are Newton's motion equation and
connected with the Newton's second law for $MP$. But in a result
of construction of the motion equation for $SP$ basing on the
Newton's laws for $MP$ it was found that the Newton's second law
is not fulfilled for $SP$. It is because the work of external
forces going not only to the motion of $SP$ but also change its
internal energy. It is the main difference between the dynamics of
the $SP$ and $MP$.

This result do not contradict to the classical mechanics. Indeed,
recall that the principle of the least of action follows from the
Newton's equation if one requires the forces acting on the system
to be conservative. Rigorous proof of this hypothesis has not been
found. So this condition was assumed a priori [3]. But it was
shown that for the non-equilibrium systems this condition is
violated. The violation has a place because the internal energy
are changes when the motion of the system in an nonhomogeneous
space [7]. Therefore the non-equilibrium systems isn't the
Hamiltonian systems and the  Poincare's theorem about
reversibility does not apply to them [11].

Thus the deterministic mechanism of the irreversibility was
obtained by the strictly mathematical calculations based on the
Newton's laws for $MP$. This was achieved by describing the system
dynamics in the micro and macro variables. Behind the simplicity
of the calculations is hidden deep physical reason the
break-symmetry of the time for the $SP$ dynamics in the
inhomogeneous space.

Initially, we note that in $LSC$ the coordinates and velocities of
each $MP$ into $SP$ are coupled and interdependent. This was shown
on the example of system of two $MP$. It is means that we have no
right to do any conclusion about the type of symmetry of the $SP$
motion equations  before transforming the dependent variables in
$LSC$ into the independent variables in the another coordinate of
the system.

The differential relations between the coordinates and velocities
of $MP$ are not differentiable in $LSC$. That is, the presence of
$MP$ interactions due to, for example, Coulombian forces, is
equivalent to the nonholonomic constraints. Therefore they do not
lead to a decrease in the dimension of the configuration space
which determines by the position of the $SP$ [8]. The nonholonomic
constraints are equivalent to the violation of potentiality of the
collective forces acting on the system [8]. It is easy to show on
the example of the task of two interacting systems [6]. Therefore
non-equilibrium systems consisting from the potential interaction
of $MP$ are not Hamiltonian's systems.

As soon as we go to the micro and macro variables, the space
variables for $SP$ are splitted into two orthogonal subspaces. In
one of them the dynamics of $MP$'s relative to $CM$ is defined
while in the second subspaces the motion of the system as a whole
is defined. From here we obtain the energy conservation law for
the $SP$. According to this law only total energy which equal to
the sum of the energy of motion and internal energy is preserved
while each of its components can not be stored in the
non-homogeneous space. I.e. the $SP$ can have different values for
the internal energy and the motion energy in the given point in
the S-space.

The change of the internal energy is determined by the nonlinear
terms of the external forces which cause an irreversible
transformation of kinetic energy of the $SP$ into internal energy.
The irreversibility connected with the inability to change the
momentum of $SP$ due to relative moves of the $MP$. It is the
essence of deterministic irreversible.

The non-equilibrium system can be submitted as a set of $SP$. Thus
in the non-equilibrium systems which are a set of $SP$, the energy
of relative motion of $SP$ is transformed into internal energy of
the $SP$. When the non-equilibrium system comes close enough to
equilibrium, the relative motions of the $SP$ disappears and then
the work on changing of the internal energy of the $SP$ disappears
too. As a result the system becomes a Hamiltonian and reversible.

Thus the nature of breaking-symmetry of the time for $SP$ associated
with the presence of internal degrees of freedom and due to
non-linear transformation of the motion energy into internal
energy. Such transformation is possible in the presence of
inhomogeneities in the external field of forces. The nonlinear
terms in the $SP$ motion equations leads to breakdown of the
invariance of the two types of energy. But the total energy $SP$
remains invariant.

In non-equilibrium systems which represented by a set of $SP$, the
irreversibility appeared due to the fact that the energy of motion
of $SP$ is transformed into internal energy of $SP$. We call this
irreversibility as a deterministic irreversibility since it
strictly follows from the Newton's laws for $MP$.

In contrast to the probabilistic mechanism of irreversibility the
deterministic irreversibility does not require a hypothesis about
the "coarse-grain" of the phase space [11]. But in the rest these
two explanations are not mutually exclusive.

Thus the irreversibility of the system's dynamics connected with
the presence of the body internal energy and the possibility of
its change in inhomogeneous external field of force. The internal
energy caused by such a class of motion of the elements which does
not change the energy of motion of the system. This class of the
motion is determined by the second term in the right-hand side of
the eq. (4).

The existence of the internal energy is possible to take into
account only through the introduction of micro variables,
determining the position and velocity elements with respect to the
$CM$ of $SP$. This means that the dynamics of systems can be
determined through the two type of variables. There are micro
variables and macro variables. The body motion is determined by
the total energy flow between the body and its environment.
Therefore it is impossible to describe the dynamics of the body
and creation a new structure if do not take into account the
change of two type of its energy.

\section{CONCLUSION}
The model of the body in the form of a set of $SP$ is more general
and closer to reality than the model of a structureless body.
Indeed, the energy of the external field does not go only to
change of the body velocity. This energy also go to the change of
the body internal energy. Therefore the dynamics of the body is
determined by the changes in the two types of energy: the internal
energy and the motion energy of the body while maintaining their
sum.

The $SP$ acceleration is not uniquely associated with the flow of
external energy as it takes place in the case for $MP$ because
part of this energy going to the increasing of internal energy.
The body acceleration is proportional to the external force only
in special cases when the internal energy of the body does not
change.

The $SP$ motion equation was obtained based on Newton's laws for
$MP$ by using the law of energy conservation. By differentiating
the $SP$ energy with respect to the time, the expression for the
energy change is defined. From this equation the $SP$ motion
equation is obtained. The $SP$ motion equation in contrast to
Newton's equation for the $MP$ includes terms that determine the
change of the $SP$ internal energy.

Derivation of $SP$ motion equations is performed in the system of
coordinates micro and macro variables taking into account that the
motion of the $SP$ in the space defined by its $CM$ and the motion
of $MP$ defined relative to the $CM$. In these variables the $SP$
energy is naturally splitted into internal energy and the energy
of motion of the $SP$. The internal energy is expressed through
the micro variability. It is determined by motion of the $MP$ in
relative to the $CM$. The energy of motion of the system expressed
in terms of macro parameters. It is energy of $SP$ motion in
space. This system of coordinate connected with the symmetry of
the systems' dynamics which splits on the symmetry of the system
and the symmetry of the space.

In contrast to the Newton's equations for $MP$, the $SP$ motion
equation is irreversible. The breaking-symmetry of the time for
$SP$ takes place because the motion energy of $SP$ transformed
into the internal energy and the internal energy can't go back to
the $SP$ motion energy. Therefor we can say that the internal
energy is the energy of the chaos.

Thus the mechanism of deterministic irreversible can be find only
when to take into account two hierarchical levels: the micro and
macro variables and the presence of the nonlinear terms in the
external forces which depending from the micro and macro
variables.

Scope of $SP$ motion equation is much broader than $MP$ motion
equation because it takes into account the energy dissipation
which is connected with the internal motions of the elements of
bodies. For example the motion equation of $SP$ in contrast to the
motion equations of $MP$ allows to describe the processes of
emergence and evolution of structures. It is because the
non-equilibrium system when the condition of local thermodynamic
equilibrium have a place, can be represented as a set of
equilibrium subsystems [10, 12] whose dynamics is described by the
$SP$ motion equation.

\smallskip

\end{document}